\documentclass[9pt, conference]{IEEEtran}
\IEEEoverridecommandlockouts
\usepackage{cite}
\usepackage{amsmath,amssymb,amsfonts}
\usepackage{graphicx}
\usepackage{textcomp}
\usepackage{xcolor}
\def\BibTeX{{\rm B\kern-.05em{\sc i\kern-.025em b}\kern-.08em
    T\kern-.1667em\lower.7ex\hbox{E}\kern-.125emX}}
\usepackage{color}
\usepackage{comment}
\usepackage[hyphens]{url}  
\usepackage{graphicx} 
\usepackage{graphicx}  
\usepackage{paralist}
\usepackage[boxed]{algorithm}
\usepackage[noend]{algpseudocode}
\usepackage{amsthm}
\usepackage{soul}
\usepackage{subfigure}
\usepackage{booktabs}
\theoremstyle{definition}
\newtheorem{definition}{Definition}[section]

\title{Intelligent Orchestration of ADAS  Pipelines on Next Generation Automotive Platforms}
\begin{document}

\author{\IEEEauthorblockN{Anirban Ghose,
Srijeeta Maity, Arijit Kar, Kaustubh Maloo, Soumyajit Dey
}
\IEEEauthorblockA{Indian Institute of Technology,  Kharagpur\\
Email: anirban.ghose@cse.iitkgp.ernet.in,\{srijeeta.maity, arijit.kar14, kaustubh.maloo\}@iitkgp.ac.in, soumya@cse.iitkgp.ac.in
}}

\maketitle
\begin{abstract}
Advanced Driver-Assistance Systems (ADAS) is one of the primary drivers behind increasing levels of autonomy, driving comfort in this age of connected mobility. However, the performance of such systems is a function of execution rate which demands on-board platform-level support. With GPGPU platforms making their way into automobiles, there exists an opportunity to adaptively support high execution rates for ADAS tasks by exploiting architectural heterogeneity, keeping in mind thermal reliability and long-term platform aging. We propose a future-proof, learning-based adaptive scheduling framework that leverages Reinforcement Learning to discover suitable scenario based task-mapping decisions for accommodating increased task-level throughput requirements
\end{abstract}
\begin{IEEEkeywords}
ADAS, OpenCL, Machine Learning, Control Theory, Heterogeneous Multicore, Real Time Scheduling
\end{IEEEkeywords}
\section{Introduction}
Recent versions of automotive software standards like {\em adaptive} AUTOSAR  \cite{autosar} recommend that multiple software features should share compute platforms in an adaptive co-scheduled manner accommodating dynamic mapping and scheduling of software tasks/features. In the context of Advanced Driver-Assistance (ADAS) software, while existing works partially explore this promise \cite{yang2019re}, they do not address how  ADAS functionalities can be mapped {\em dynamically} on modern multicore  platforms which are also heterogeneous in nature, i.e. different cores on the SoC adhere to different computing paradigms like control-flow  intensive processing of CPUs, SIMD style high throughput processing of GPUs, re-configurable blocks like FPGAs etc. 
Accommodating dynamic task mapping requests  can be  difficult  in heterogeneous architectures due to the following issues, - i) the suitability of a specific task on a core type needs to be learned using profiling runs, ii) a  mapping request needs to be satisfied while keeping in mind the  architectural demands of other existing tasks.  
\par 
An ADAS system constitutes 
multiple object detection pipelines that process sensor data periodically leveraging state of the art Deep Neural Networks (DNNs) and Convolutional Neural Networks (CNNs). The pipelines are used to detect objects in the vicinity and accordingly dispatch commands to other vehicular subsystems such as park assist, anti-lock braking systems etc. for taking relevant actions. Therefore, there exists a natural requirement for real time guarantees for executing these object detection pipelines. Recent works \cite{zhou2018s,yang2019re} emphasize on designing efficient scheduling algorithms at the system level in addition to algorithmic optimizations on neural network workloads \cite{ren2015faster,redmon2016you} for meeting these real time requirements. Additionally, ADAS detection pipelines impose different frames per second (FPS) requirements for different detection tasks depending on the current environmental context. Even for the same detection task, given different driving scenarios, these FPS requirements are subject to change to meet a desired level of object detection accuracy. For example, a pedestrian detection system would like to process images at a higher frame rate if the on-board GPS points to the fact that the vehicle is approaching a congested area. In situations like this, for accommodating increased FPS requests, the underlying scheduler must allocate resources for increased object detection accuracy while maintaining real time guarantees. 
The state of the art ADAS scheduling algorithms for heterogeneous CPU-GPU platforms optimize the  execution of ADAS pipelines on the GPU by i) decreasing the overall latency of detection jobs via software pipelining approaches so as to increase the object detection accuracy \cite{yang2019re}, ii) leveraging algorithmic fusion techniques for processing multiple frames concurrently \cite{yang2019re} or iii) opting for data fusion approaches followed by concurrent execution of multiple DNN workloads in the GPU hardware \cite{zhou2018s}. We note that the existing approaches are not equipped to handle tasks with time varying dynamic FPS requirements dictated by different driving contexts. Additionally, leveraging a single line of approach for scheduling these pipelines may not always yield the best possible scheduling solution. For example, fusing and processing frames concurrently on hardware impose a high memory footprint for selected pipelines and may not be a feasible approach always. Given the set of ADAS detection workloads, designing automotive computing  solutions with the ability to sustain the maximum FPS requirements of all detection pipelines simultaneously may lead to over-provisioning of resources on a restricted memory architecture, and also increased power consumption and thermal aging of the heterogeneous platform. A generalized approach  exploring opportunistic GPGPU optimizations must be envisioned that combines the techniques mentioned above for ascertaining optimal application to architecture scheduling decisions at runtime while keeping in mind the overall power budget of the target platform. Since investigating every possible mapping decision imposes a considerably large search space of decisions to be evaluated for finding an optimal solution, an intelligent task manager needs to be designed which will predict task-device mapping decisions for each pipeline subject to dynamic FPS requirements over time.
\begin{figure}[ht]  
	\centering
	\includegraphics[scale=0.47]{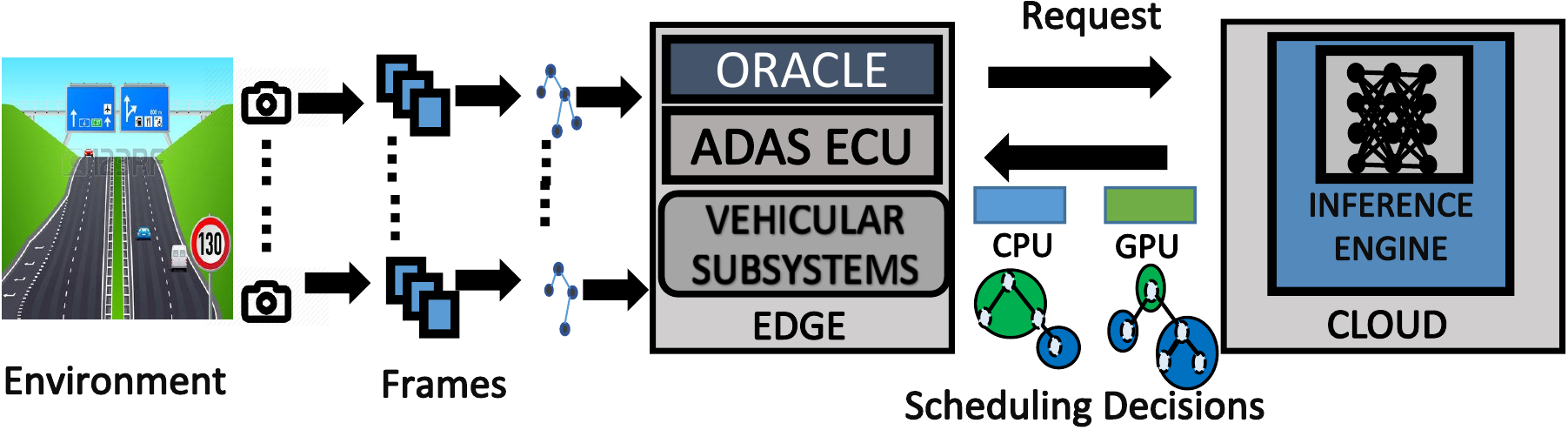}
	\vspace{-1mm}
	\caption{Runtime System Overview\label{fig:overview}}
\end{figure}
\vspace{-1mm}
\par The present work proposes an intelligent runtime system which can manage the mapping and scheduling of ADAS detection pipelines in next-generation automotive embedded platforms in a self-learning fashion so that the time varying dynamic detection requirements of existing pipelines are efficiently managed while maintaining real time guarantees. 
An overview of the proposed software architecture for our runtime system is depicted in Fig. \ref{fig:overview}.
We assume there exists an  oracle executing as a service on the vehicle software stack which keeps track of environmental parameters such as the terrain the vehicle is currently driving in, observes the output of existing ADAS detection pipelines, interacts with on-board  sensors (e.g. GPS) and generates requests that characterize FPS requirements for detection job(s). 

\par Our proposed intelligent runtime system comprises an inference engine which exist as a service running on a cloud server. The engine leverages a learned model trained using Reinforcement Learning (RL) techniques. The model is specific for the current set of detection pipelines executing on the ADAS ECU and is periodically retrained in the cloud whenever an over-the-air software update occurs such as i) the injection of a new detection pipeline in the current set of jobs and ii) the refinement of parameters for an existing pipeline. 
The inference engine uses the learned model to determine a set of task-device mapping decisions and accordingly informs the oracle whether the requests can be accommodated or not. 
\par The admissible decisions are communicated to a low level scheduler running on the ADAS ECU. 
The decisions reported by the inference engine are based on ground-truth learned models which assume that latency of a task for a predicted mapping decision remains constant for all invocations on a given device. However, this is not the case for a shared compute platform due to interference from other tasks. For handling such mispredicted scenarios,  the low-level scheduler employs  state-of-the-art  control-theoretic scheduling approaches that apply relevant core level DVFS to ensure predictable task latency. 
The salient features of the proposed work are summarized as follows.
\begin{compactitem}
\item We characterize a Reinforcement Learning (RL) based problem formulation in the context of real time scheduling on ADAS platforms and present a training methodology for the same.
\item We present an inference engine which is a discrete event scheduling simulator that 
determines task-device mapping decisions subject to dynamic oracle requests.
\item We create an intelligent runtime scheduler which leverages control-theoretic schemes to mitigate potential deadline misses due to bad quality mapping decisions. 
We provide extensive validation results justifying the usefulness of our RL assisted ADAS deployment architecture. 
\end{compactitem}
\section{Problem Formulation}
Let  $\mathcal{J} = \lbrace G_1, G_2 , \cdots ,G_N\rbrace$ be the set of $N$ ADAS jobs to be scheduled. We model an ADAS job as a directed acyclic graph (DAG) $G_k = \langle T_k,E_k \rangle $ where $T_k = \lbrace t^k_1, t^k_2,\cdots,t^k_n\rbrace$ denotes the set of tasks, $E_k \subseteq T_k \times T_k$ denotes the set of edges where each edge $(t^k_i, t^k_j)$ denotes that task $t^k_j$ cannot start execution until and unless $t^k_i$ has finished for a DAG $G_k$. In the context of ADAS detection pipelines, each task refers to a data parallel computational kernel \cite{stone2010opencl}. Given $\mathcal{J}$, we denote an oracle request to be of the form $\mathcal{R}$ = $\lbrace \langle G_1,w_1,p_1 \rangle , \langle G_2,w_2,p_2 \rangle, \cdots, \langle G_N,w_N,p_N \rangle\rbrace$ where each tuple here specifies that every job $G_k \in \mathcal{J}$ arrives periodically with period $p_i$ and processes $w_i$ frames in each execution instance.  The set of requests $\mathcal{R}$ is made by the oracle based on observations of the current driving scenario. As long as the scenario does not change, the request is maintained at its current state. Note that existing ideas of processing multiple frames concurrently at a given rate as well as fusing the decision over multiple frames by increasing the execution frequency, can both be captured by our task models. 
Given the specification for each DAG $G_k$ in $\mathcal{R}$, let us denote $\mathcal{G}_k = \{G^{1}_k,\cdots,G^h_k \}$ as the set of all execution instances of DAG $G_k$ where $h = H/p_k$ and $H$ is the hyper-period which is the l.c.m of the periods (inverse of rate) of DAGs. We denote the $j^{th}$ execution instance of a DAG $G_k$ as $G^j_k$ and denote the $i^{th}$ task belonging to it as $t^{j,k}_i$. 
We define the set $\mathcal{G} = \mathcal{G}_1 \cup \mathcal{G}_2 \cup \cdots \cup \mathcal{G}_N$ to be the  set of all execution instances of all DAGs in the job set $\mathcal{J}$ executing in the hyper-period $H$. Given $\mathcal{G}$, we denote a hyper-period snapshot $\mathcal{H}$ to be the set of tasks of each DAG execution instance $G^j_k \in \mathcal{G}$ waiting for a dispatch decision. 
We say that every DAG $G^j_k \in \mathcal{G}$ has finished execution {\em iff} $\mathcal{H}$ becomes empty. We denote $\mathcal{F}(\mathcal{H})$ as the head of $\mathcal{H}$ comprising  tasks that are ready to execute and define it as follows. 
\begin{definition}
\par \noindent Given a hyper-period snapshot $\mathcal{H}$, the \textbf{frontier} $\mathcal{F}(\mathcal{H})$ is typically a set of independent tasks belonging to a subset of DAGs $\mathcal{G}^\prime \subseteq \mathcal{G}$ such that the following precedence constraints hold: i) each DAG $G^j_k = \langle T^j_k,E^j_k \rangle \in \mathcal{G}^\prime$ must finish execution before $G^{j+1}_k$ and ii) for each DAG $G^j_k \in \mathcal{G}^\prime$, predecessors of each task $t^{j,k}_i\in T^j_k$ belonging to $\mathcal{F}(\mathcal{H})$ i.e tasks $t^{j,k}_l$ such that $(t^{j,k}_l, t^{j,k}_i) \in E^j_k$ have finished execution.

\noindent The frontier $\mathcal{F}$ is ordered by the following ranking measures.
\end{definition}
\begin{definition}
The rank measure \textbf{blevel} of a task $t$ in DAG $G$=$\langle T,E \rangle$ represents the best case execution time estimate to finish tasks in the longest path starting from $t$ to a task that has no successors in $G$ assuming all resources are available and is computed as  $blevel(t)$=$e_t$+$\max_{t' \in succ(t)} blevel(t')$, where $e_t$ is the worst case execution time (WCET) of task $t$ and $succ(t)=\lbrace t^\prime | (t,t^\prime) \in E \rbrace$.
\end{definition}
\begin{definition}
The rank measure \textbf{local deadline} of a task $t$ in DAG instance $G$=$\langle T,E \rangle$ represents the absolute deadline of $t$ and is computed as $local\_deadline = d-blevel(t)+e_t$, where $d$ is the absolute deadline of $G$, $blevel$ is the aforementioned rank measure and $e_t$ is the WCET of $t$. 
\end{definition}
\par\noindent\textbf{Task execution flow:} Let us consider for some hyper-period snapshot $\mathcal{H}$, a frontier of tasks $\mathcal{F}(\mathcal{H})$ sorted by the {\em local deadline} ranking measure.
Given this sorted list of tasks in  $\mathcal{F}(\mathcal{H})$, the set of available devices in a target heterogeneous multicore $\mathcal{P}$, 
and the task $t_{min} \in \mathcal{F}(\mathcal{H})$ with the minimum {\em local deadline} as inputs, let a mapping function $\mathcal{M}$ return a task-device mapping $m= \langle T,P \rangle$ where $T$ comprises a set of tasks comprising the task $t_{min}$ and its descendants. The quantity $P$ represents one device in the target heterogeneous platform $\mathcal{P}$. The choice of $T$ is motivated by the fact that in certain runtime contexts it may be beneficial that multiple tasks/kernels are fused and mapped to $P$ for achieving better register and  cache usage and avoiding the  launch overhead of individual tasks. Related works \cite{qiao2018automatic,xing2019dnnvm} have been proposed over the years which investigate the efficacy of kernel fusion on heterogeneous CPU/GPU architectures by considering different runtime contexts. The primary objective of this work lies in learning these contexts in the form of a policy function $\pi$ which may be used to design $\mathcal{M}$ that modifies the hyper-period states. The reason for leveraging a learning based approach can be attributed to the large space of scheduling decisions that are possible using kernel fusion. Considering a DAG $G$ of depth $D$, we assume only vertical fusion i.e. all tasks selected upto a particular depth in the mapping decision $m=\langle T,P \rangle$ are fused and mapped to $P$. The total number of possible fusion based mapping configurations for the entire DAG $G$ is therefore equal to the number of integer compositions of $D$ which is $2^D$ \cite{Opdyke2010}. Furthermore, since each fusion based mapping configuration has the option of getting mapped to a CPU or GPU device, the total number of possible scheduling decisions for a single DAG is actually $\omega (2^{D})$. Considering $m$ jobs in $\mathcal{J}$, the total space of scheduling decisions is $\omega (2^{mD})$. We next elaborate with an illustrative example how $\mathcal{M}$ is used to obtain mapping decisions in this exponential search space.
\par 
\begin{figure}[ht]  
	\centering
	\includegraphics[scale=0.52]{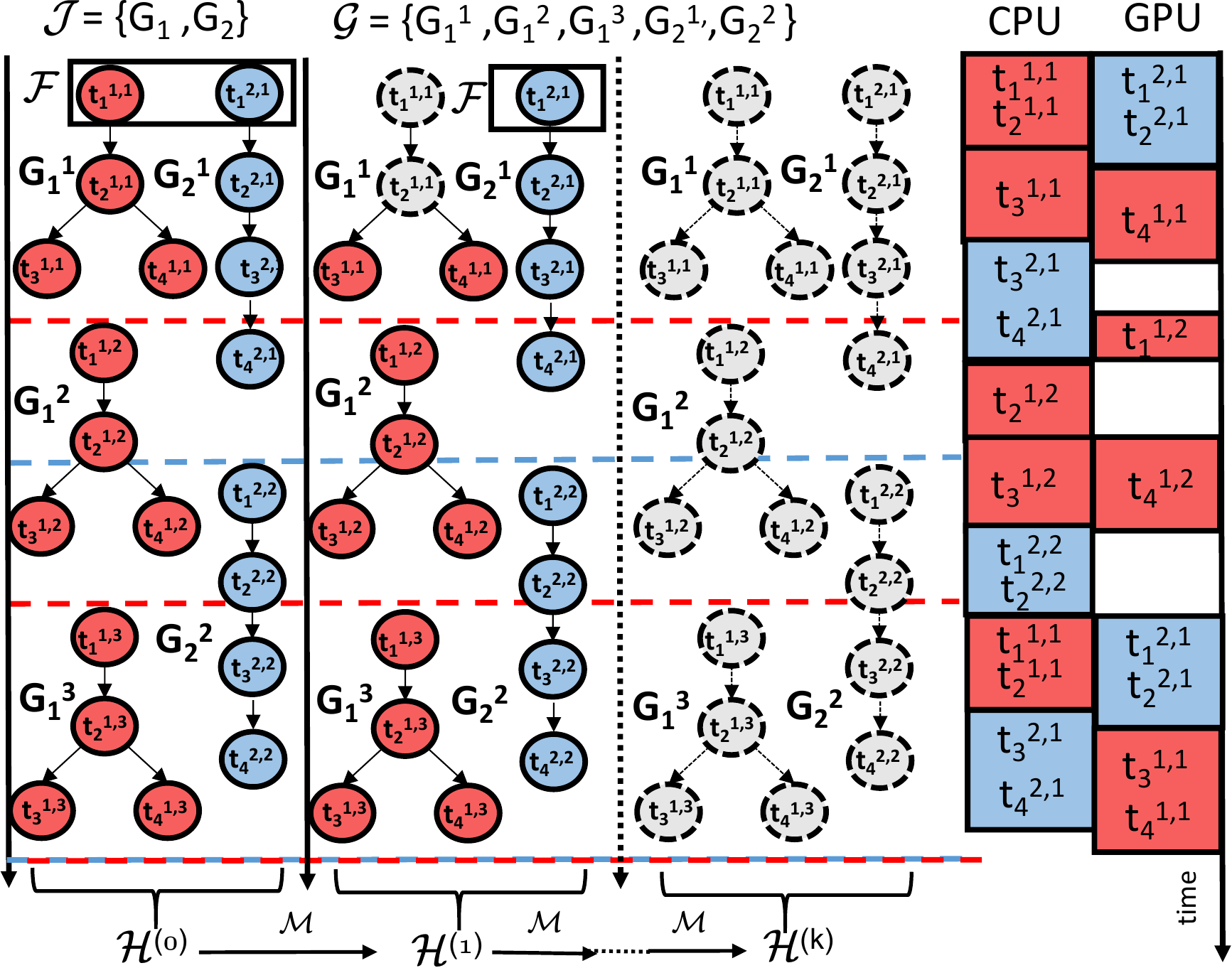}
	\caption{RL Assisted Task Mapping\label{fig:probform}}
\end{figure}
\par Given $t_{min} \in \mathcal{F}(\mathcal{H})$,
the function $\mathcal{M}$ is applied on $\mathcal{H}$ to yield a new hyper-period snapshot $\mathcal{H}^\prime$ and a new frontier $\mathcal{F}(\mathcal{H^\prime})$. 
This process of applying $\mathcal{M}$ 
and generating subsequent hyper-period snapshots is continued until each DAG in $\mathcal{G}$ has been scheduled. We present a representative example depicting a complete schedule of task-device mappings for a job set $\mathcal{J}= \{G_1,G_2\}$ in Fig. \ref{fig:probform}. 
\par \noindent Given a hyper-period $H$, there exists three instances of job $G_1$ (deadlines are shown using dashed red lines) and two instances of job $G_2$ (deadlines are shown using dotted blue lines). The set $\mathcal{G}$ is therefore $\{ G^1_1,G^2_1,G^3_1,G^1_2,G^2_2\}$.  Initially the frontier contains tasks $G^1_1$ and $G^1_2$ which have no predecessors  i.e. $\mathcal{F} = \{t^{1,1}_1, t^{2,1}_1 \}$. Tasks are selected from $\mathcal{F}$ based on the local deadline rank measure. 
The mapping function $\mathcal{M}$ generates a task-device mapping decision $m = \langle \lbrace t_1^{1,1}, t_2^{1,1} \rbrace, CPU \rangle$. This is shown in the Gantt chart in the right hand side of Fig. \ref{fig:probform}. This in turn creates the hyper-period snapshot $\mathcal{H}^{(1)}$ as depicted in the figure. 
 The corresponding list of mapping decisions is depicted in the Gantt chart schedule in Fig. \ref{fig:probform}. We summarize our formulation of learning based dispatch as follows.
 \par {\em Given a set $\mathcal{G}$ constructed from a job set $\mathcal{J}$ of ADAS detection pipelines for a given oracle request $\mathcal{R}$ to be executed on the heterogeneous platform $\mathcal{P}$, the objective of the proposed scheduling scheme is to leverage a policy function $\pi$ trained using RL which dictates the choice of task-device mapping function $\mathcal{M}$.} 
 \par The set of task-mapping decisions reported by $\pi$ are platform agnostic and does not consider the latency variation that might occur due to platform level interference factors such as thermal throttling, memory thrashing, shared memory contention by other tasks etc. This variation may potentially lead to scenarios where the actual deadline miss rate exceeds the predicted deadline miss rate. For this purpose, in the deployment phase, the scheduling scheme monitors the execution times of each task-device mapping decision and enters into a safe control-theoretic mode of dispatch whenever it observes a potential deadline violation. To understand this we define the following measure.
 \begin{figure}[ht]  
	\centering
	\includegraphics[scale=0.5]{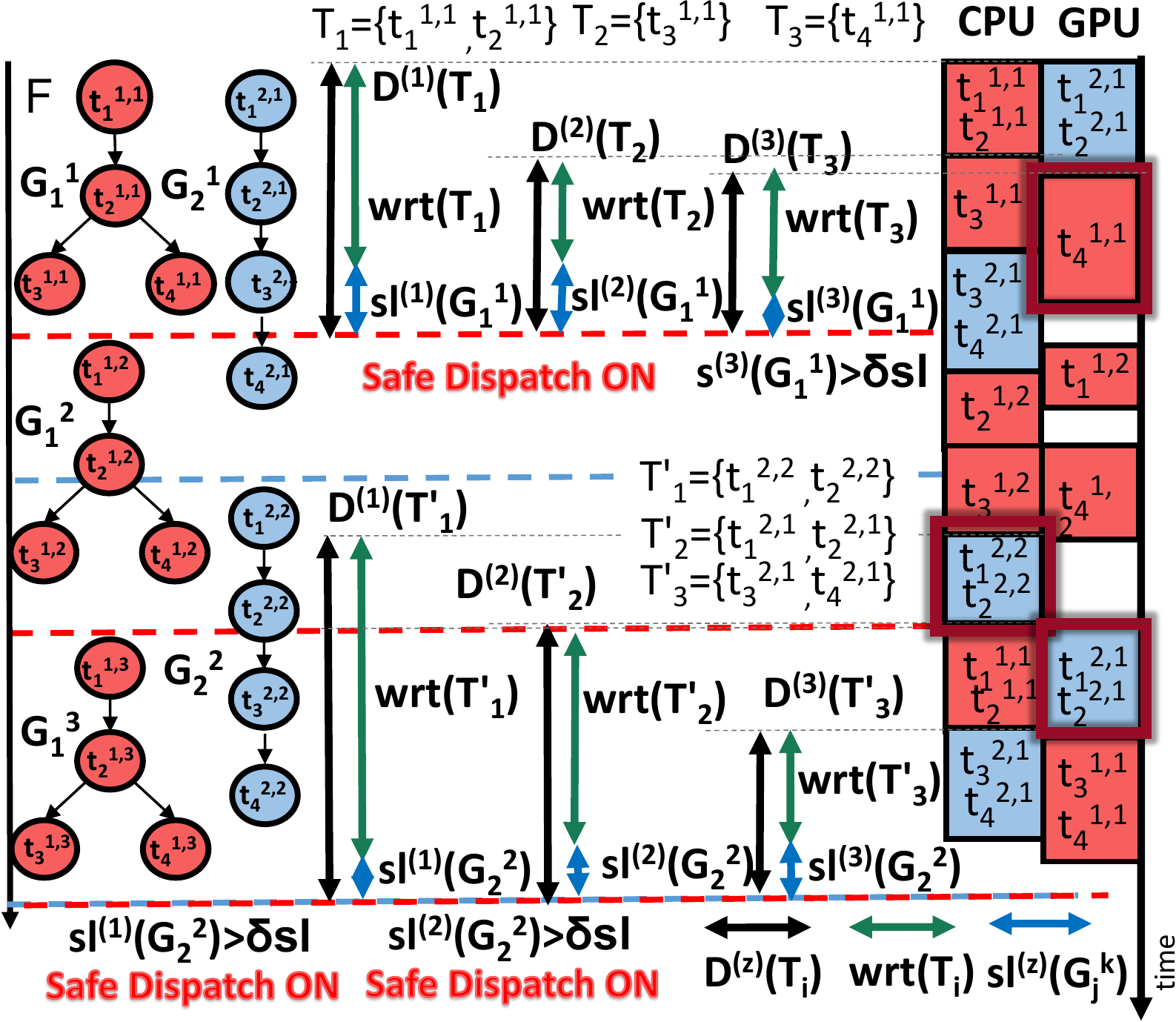}
	\caption{Safe Low Level Scheduling \label{fig:slack}}
\end{figure}
 \begin{definition}
 The slack measure for the  $z^{th}$ task mapping decision $m =\langle T,P\rangle$ pertaining to some DAG instance $G^j_k$ is defined as $sl^{(z)}(G^j_k)$ = $D^{(z)}(T)- wrt(T)$ where $D^{(z)}(T)$ denotes the current time remaining to meet the deadline for the DAG instance $G^j_k$ and $wrt(T)=\sum_{t\in T \cup SUCC(T)}(e_t)$ denotes the worst case remaining time for executing tasks belonging to $T$ and the set $SUCC(T)$ which comprises all descendants of tasks in $T$ in $G^j_k$. 
 \end{definition}
 In order to ensure that deadline requirements are met at runtime, the inequality $sl^{(z)}(G^j_k) \geq \delta sl$ must be respected, where we use $\delta sl$ to model the overall uncertainty associated with the WCET estimates of the tasks constituting the DAG. The scheduling scheme enters into the safe mode if it observes that $sl^{(z)}(G^j_k) < \delta sl$ and applies core-level DVFS iteratively to each successive mapping decision for $G^j_k$ until it is ensured that $sl^{(z')}(G^j_k) \geq \delta sl$ where $z'>z $. 
 \par We elaborate the safe low-level dispatch mechanism with the help of Fig. \ref{fig:slack}. The sequence of mapping decisions for $G^1_1$ is the set $\{ \langle T_1,CPU \rangle, \langle T_2,CPU \rangle, \langle T_3,GPU \rangle \}$. We observe for $T_2$, that $sl^{(2)}(G_1^1) \geq \delta sl$ and thus the safe mode of dispatch for the scheduler is not engaged. However for $T_3$, we observe that $sl^{(3)}(G_1^1) < \delta sl$. Even though $T_1$ had finished execution, $T_3$ could not start immediately because the GPU device was engaged by tasks in DAG $G^1_2$. This in turn affected the deadline requirement $D^{(3)}(T_3)$ for $T_3$, initiating the safe mode of dispatch, which increased the frequency of the GPU device to ensure that $G^1_1$ respected its deadline. Similarly, considering the set of mapping decisions for $G^2_2$ i.e. $\{ \langle T'_1,CPU \rangle, \langle T'_2,GPU \rangle, \langle T'_3,CPU \rangle \}$, it may be observed in Fig \ref{fig:slack}, that the scheduler engages into safe mode for $T'_1$ and $T'_2$ and applies core-level DVFS for the CPU and GPU devices respectively. The delay in starting $T'_1$ due to resource contention of both CPU and GPU devices by tasks in $G^2_1$ violates the slack constraints and forces the safe mode to be initiated. 
 \par The proposed system-level solution for real time ADAS scheduling in this communication therefore operates in two distinct phases -i) an AI enabled approach which searches through the exponential space of scheduling decisions and intelligently selects a global set of task-mapping decisions for each ADAS pipeline and ii) a control-theoretic scheme which performs locally for a particular pipeline. The first phase extracts from the exponential search space, the possibly best global  scheduling decisions. The second phase is initiated for these decisions only if the runtime slack constraints defined above are violated. The combined approach therefore ensures opportunistic switching to high frequency mode thus reducing thermal induced degradation of lifetime reliability for the overall platform; something which can happen with pure frequency scaling based task scheduling techniques.  

\vspace{-3mm}
\section{Methodology}
\par In the recent past, several works \cite{mao2016resource,fang2017qos,domeniconi2019cush} have emerged which leverage deep reinforcement learning methods to learn an optimal policy function for solving scheduling problems. We leverage sample efficient RL approaches such as {\em Q-learning} with experience replay for training DQNs as well as Double DQNs (DDQNs) for learning state-action value functions $\mathcal{Q}(s,a)$ that characterize the goodness of choosing an action $a$ given a state $s$.  
\begin{figure*}[ht] 
\centering
\includegraphics[height=2.25in, width=7in]{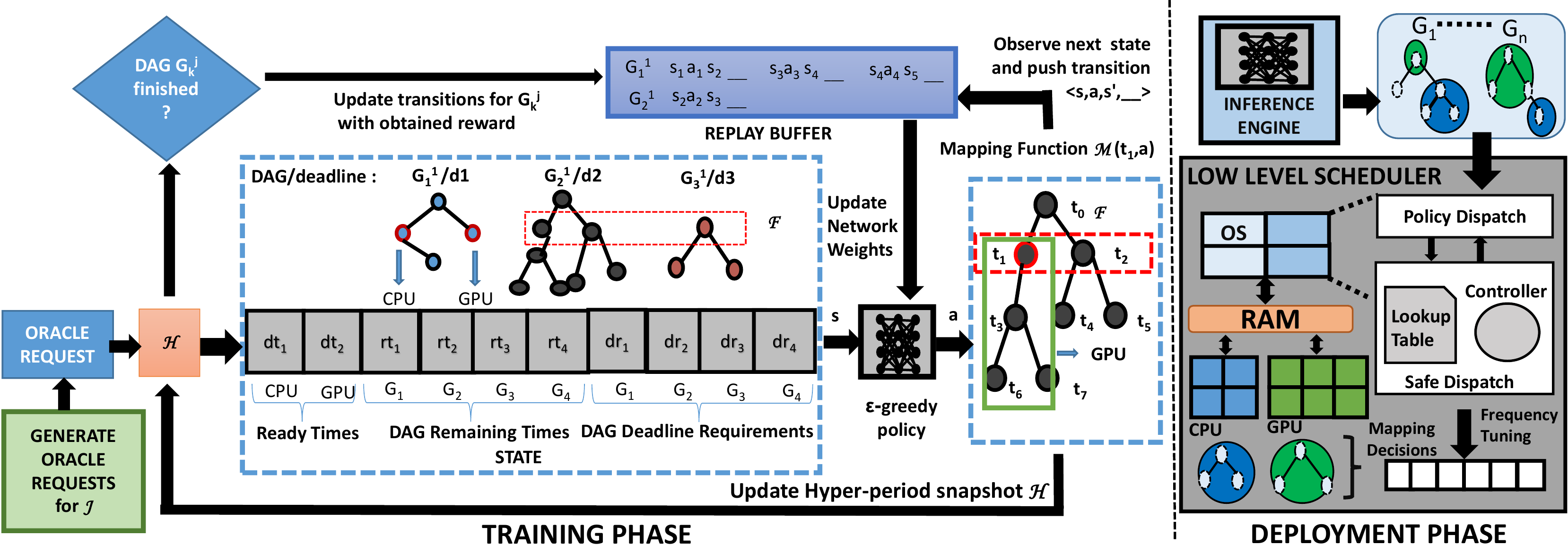}
\caption{Training and Inference Methodology Overview\label{fig:rlmethodology}}
\end{figure*}
The overall methodology depicting the training process for ascertaining $\mathcal{Q}(s,a)$, followed by  subsequent usage of the trained model in the deployment phase is next elaborated with the help of Fig. \ref{fig:rlmethodology}.
\par \noindent \textbf{Training Phase: }The input to the training phase is a set of oracle requests pertaining to the set of ADAS jobs $\mathcal{J}$ and the set of target platform devices $\mathcal{P}$. The output is a state-action value function $\mathcal{Q}$. For each such oracle request, the training process involves a series of steps for updating the network weights of $\mathcal{Q}(s,a)$. This is explained as follows. 
\par \noindent \textbf{(i) State Extraction:} Given an oracle request $\mathcal{R}$, the hyper-period snapshot $\mathcal{H}$ is first constructed. 
The observed state vector $s$ for $\mathcal{H}$ at any point of time is defined as a vector $[dt_1,dt_2,\cdots,dt_{|\mathcal{P}|},rt_1,rt_2,\cdots,rt_{|\mathcal{J}|},dr_1,dr_2,\cdots,dr_{|\mathcal{J}|}]$  where $dt_i$ represents the time  left  for the $i^{th}$ device $\in \mathcal{P}$ to become free. The quantity $rt_i$ denotes the `best case' time estimate remaining for the currently executing instance of DAG in $G_i =\langle T_i,E_i \rangle\in \mathcal{J}$ to finish. This is calculated by the $blevel$ estimate of the current task $t \in T_i \in  \mathcal{F}$. The quantity $dr_i$ represents the `time to deadline' estimate for the same instance of the $i^{th}$ DAG in $\mathcal{J}$ to finish so that the deadline of $G_i$ is respected. This is obtained by the difference between the time elapsed since the beginning of the hyper-period and the absolute deadline of the DAG instance $G_i$ in that hyper-period. The three quantities defined together capture resource availability ($dt_i$), estimate of the `best case' time  for a DAG to finish completely ($rt_i)$, and estimate of the time in which the DAG must finish for respecting deadline constraints (determined by $dr_i$). In Fig. \ref{fig:rlmethodology}, considering a system with 1 CPU and 1 GPU device and a total of 4 periodic DAGs, a sample state vector is depicted.
\par \noindent \textbf{(ii) Action Selection:} Given the observed state $s$ as input, the action $a$ is determined using an $\epsilon-$greedy policy function. Given the action $a$ and the task with the earliest local deadline $t_{min} \in \mathcal{F}(\mathcal{H})$, we define the mapping function $\mathcal{M}(t_{min},a) =\langle T,P\rangle$ where $T$ is the set of tasks comprising $t_{min}$ and its descendants up to a depth $d$ and a device $P \in \mathcal{P}$. In our setting, the value of the action $a$ can be used to infer both $d$ and $P$ as follows. An action $a$ is represented as an integer $a \in[0,\cdots,n(D+1)-1]$ where $D$ denotes the maximum height of a DAG $G_i \in \mathcal{J}$ and $|\mathcal{P}|=n$ is the total number of devices/cores in the target platform. 
An action value $a=i(D+1)+d$ represents fusing a task with descendants up to depth $d$ and mapping to the $i$-th device so that given $a$, we have fusion depth $d=a\%(D+1)$ and device id $i=a/(D+1)$ considering $[0,D]$ as domain of depth values and $[0,n-1]$ as domain of device ids. 
\par \noindent \textbf{(iii) Reward Assignment:} Once the mapping function $\mathcal{M}(t_{min},a)$ is applied, the current hyper-period snapshot $\mathcal{H}$ is updated and the subsequent next state $s'$ is observed. An incomplete transition tuple of the form $\langle s,a,s'\_ \rangle$ is pushed to a replay buffer memory $\mathcal{B}$ which is used during training updates.  We note that a transition tuple is pushed to $\mathcal{B}$ for each such action mapping decision for some task $t_i^{j,k}$ belonging to DAG instance $G_k^j$. The final empty element of the tuple represents the future reward $r$ for this transition which is updated once the DAG instance $G_k^j$ finishes execution.  A reward of value of $-1$ is assigned if it is observed that the finishing timestamp of $G^j_k$ exceeds its absolute deadline and a reward of $+1$ for the alternate case. The same reward value $r$ is used to update every incomplete transition tuple in $\mathcal{B}$ pertaining to each task action mapping decision taken during the lifetime of $G_k^j$.
\par \noindent \textbf{(iv) Training Updates: } Training updates are done by sampling the replay memory $\mathcal{B}$ randomly, constructing a set $B\subseteq \mathcal{B}$ of complete transition tuples and calculating the average loss function $\mathcal{L} = \frac{1}{\mathcal||B||} \left\{\sum_{(s,a,r,s') \in \mathcal{B}} L(\delta(s,a,r,s'))\right\}$ where $L$ is  the Huber Loss function \cite{huber2011robust} of the error term $\delta(s,a,r,s')$. The quantity $\delta(s,a,r,s')$ represents the temporal difference (TD) error for a transition tuple $(s,a,r,s')$. We consider two standard training paradigms - i) learning a simple Deep Q Network (DQN) where  TD error $\delta(s,a,r,s') = \mathcal{Q}(s,a) - (r+\gamma max_a \mathcal{Q}(s',a))$ and ii) learning a Double DQN (DDQN) \cite{van2016deep} approach where $\delta(s,a,r,s') = \mathcal{Q}(s,a) - (r+\gamma \mathcal{Q}(s',argmax_a\mathcal{Q}(s',a)))$. The quantity $\gamma$ represents the discount factor and is set to one given the episodic setting of our problem. The network is updated by mini-batch gradient descent using the loss calculated multiple times for each training run pertaining to a given oracle request. 
\par The overall steps discussed so far for updating the network $\mathcal{Q}$ is repeated for each oracle request $\mathcal{R}_i$, for a total of $num\_runs$  number of times which is an experimental parameter. The entire process of invoking training updates for each run of each oracle request is iteratively repeated until the average reward observed for oracle requests converge. The trained network $\mathcal{Q}$ is used to obtain the corresponding optimal policy $\pi^*(s)=argmax_a\mathcal{Q}(s,a)$ which is leveraged in the deployment phase.

\par\noindent\textbf{Deployment Phase:} Given an oracle request $\mathcal{R}_i$ and the resulting set of DAG instances $\mathcal{G}$, the inference engine considers the hyper-period snapshot $\mathcal{H}$  corresponding to $\mathcal{R}_i$ and iteratively does the following steps - i) observes state $s$, ii) uses $\pi^*(s)$ to select action $a$, iii) applies mapping function $\mathcal{M}$ using selected action $a$ on $\mathcal{H}$. The entire inference process invokes multiple inference passes over the learned network until $\mathcal{H}$ becomes empty, finally yielding a set of task-device mapping decisions $\Pi(\mathcal{G})$. Using these decisions and the available WCET estimates of tasks, the engine simulates the schedule specified in $\Pi(\mathcal{G})$, assesses the percentage of deadline misses and accordingly suggests admissible scheduling decisions to the low-level scheduler. 
\begin{algorithm}[H]
	\scriptsize
	\caption{Control Theoretic Scheduling Scheme \label{algo:safe}}
	\begin{algorithmic}[1]
	    \State $sp(z-1) \leftarrow 1$
		\For{each mapping $m_{z} \in \Pi (G_k^j)$ }		
    		\State $\langle T_z,P_z\rangle \leftarrow m_z$
    		\State $D^{(z)}(T_z) \leftarrow$ observe current time to deadline
    		\State $sl^{(z)}(G^j_k) \leftarrow  D^{(z)}(T_z)- wrt(T_z)$
    		\If{$sl^{(z)}(G^j_k) < \delta sl$}
    		    \State $sp(z)=sp(z-1)+ \rho*err(z)/b(z)$ 
    		    \State freq = lookup($sp(z),P_z$)
    		    \State set $P_z$ frequency to freq
    		\EndIf    		
    	\EndFor
	\end{algorithmic}
\end{algorithm}
\vspace{-2mm}
The low level scheduler maps tasks following decisions specified in $\Pi(\mathcal{G})$ and uses the local control theoretic scheduling  scheme outlined in Algorithm \ref{algo:safe} for each sequence of mapping decisions $\Pi(G^j_k)$ pertaining to each DAG instance $G^j_k$. 
The scheme is inspired from the state-of-the-art pole-based self-tuning control techniques \cite{Mishra:2018:CLC:3173162.3173184} that dynamically model the speedup of $T$ as a function of core clock frequencies of the device $P$. 
The algorithm iterates over each mapping decision $m_z = \langle T_z,P_z\rangle$ in $\Pi(G^j_k)$ (lines 2-9), checks the slack constraint $sl(z)(G^j_k) \geq \delta sl$ and increases the core frequency of $P_z$ for the duration of $T_z$ if required.
This is done using the speedup equation  $sp(z)=sp(z-1)+ \rho*err(z)/b(z)$ (line 7) where $sp(z)$ denotes the speedup requirement for $T_z$ such that the error term $err(z) = sl^{(z)}(G^j_k) - \delta sl$ is rendered positive. The quantity $b(z)=\sum_{t\in T}e_t $ represents the WCET estimate for executing $T$ at the baseline frequency and $\rho$ represents the pole value of the controller. Given the required speedup, the scheme uses lookup tables computed offline during the profiling phase which map speedup values of $T_z$ to core frequency values of the device $P_z$ to obtain the required operating frequency $freq$. The core-level frequency of $P_z$ is increased to $freq$ and task $T_z$ is executed. This process is repeated only for those task-device mapping instances that violate the slack constraints.

\section{Experimental Results} 
We consider the Odroid XU4 embedded heterogeneous platform comprising two quad-core ARM CPUs (Big and Little), and one Mali GPU.  We map 1) the host OS (Ubuntu 18.04 LTS) on two cores of  Little CPU , 2) our low level scheduler as an  independent OpenCL process in the other two cores of  Little CPU, 3) the ADAS detection pipelines in the Big CPU and the GPU. We leverage OpenCL \cite{stone2010opencl}, a popular heterogeneous computing language for implementing these object detection pipelines. 
For our experimental evaluation, we have implemented a total of four representative object detection pipelines from scratch where two pipelines ($G_1$ and $G_2$) represent vanilla DNN benchmark implementations, each comprising 5 tasks and the remaining two ($G_3$ and $G_4$) represent CNN benchmark implementations, each comprising 6 tasks. We have built these pipelines using platform optimized implementations of elementary data parallel kernels (such as convolution, general matrix multiplication, pooling, softmax etc. ) available in the ARM OpenCL SDK \cite{armsdk}. Our experiments require profiling data for each task as well as each fused task variant on the target platform for setting up the environment in our training phase. We have  developed a code template generator which automatically synthesizes OpenCL code for all possible fused task  variants that are possible for each pipeline ($5\times(5-1)/2=10$ for the DNN benchmarks and 15 for the CNN benchmarks). This is useful for on-the-fly fused variant generation of future pipelines which may be downloaded on the platform.
\par\noindent\textbf{Environment Setup} The WCET estimates of each task and each fused task variant in each of the pipelines are obtained by leveraging a co-run degradation based profiling approach outlined in \cite{zhu2017co}. While profiling each benchmark on a particular device (Big CPU or Mali GPU), we execute a micro-kernel benchmark continuously on the other device in parallel to ensure maximum shared memory interference on the target platform. The WCET estimates $\tau_{CPU}(t_i$) and $\tau_{GPU}(t_i$)  represent the time taken (averaged over 10 profiling runs) to execute the task $t_i$ on the  CPU or GPU device respectively in the worst possible scenario when the system memory bandwidth is completely exploited. Using these WCET estimates, the overall WCET of the DAG $G_i = \langle T_i,E_i \rangle$ is given by $\tau(G_i) = \sum_{t \in T_i} max(\tau_{CPU}(t_i),\tau_{GPU}(t_i))$. 
\par The oracle request $\mathcal{R}$ for our experiments takes the form $\lbrace \langle G_1,w_1,p_1 \rangle, \langle G_2,w_2,p_2 \rangle, \langle G_3,w_3,p_3 \rangle,\langle G_4,w_4,p_4 \rangle\rbrace$, with $w_i=1$ for all DAGs. For generating oracle requests, we vary $p_i$ for each $G_i$ with values from the set $\lbrace \tau(G_i), 2*\tau(G_i), 3*\tau(G_i) \rbrace$. Since each DAG processes one frame at a time and can arrive using one of the three period values, the total number of oracle requests possible for the job set comprising 4 DAGs is $3^4=81$. We train our DQN and DDQN using these 81 oracle requests and discuss our findings below.
\vspace{-2mm}
\begin{figure}[ht]  
\centering
\includegraphics[width=.4\textwidth]{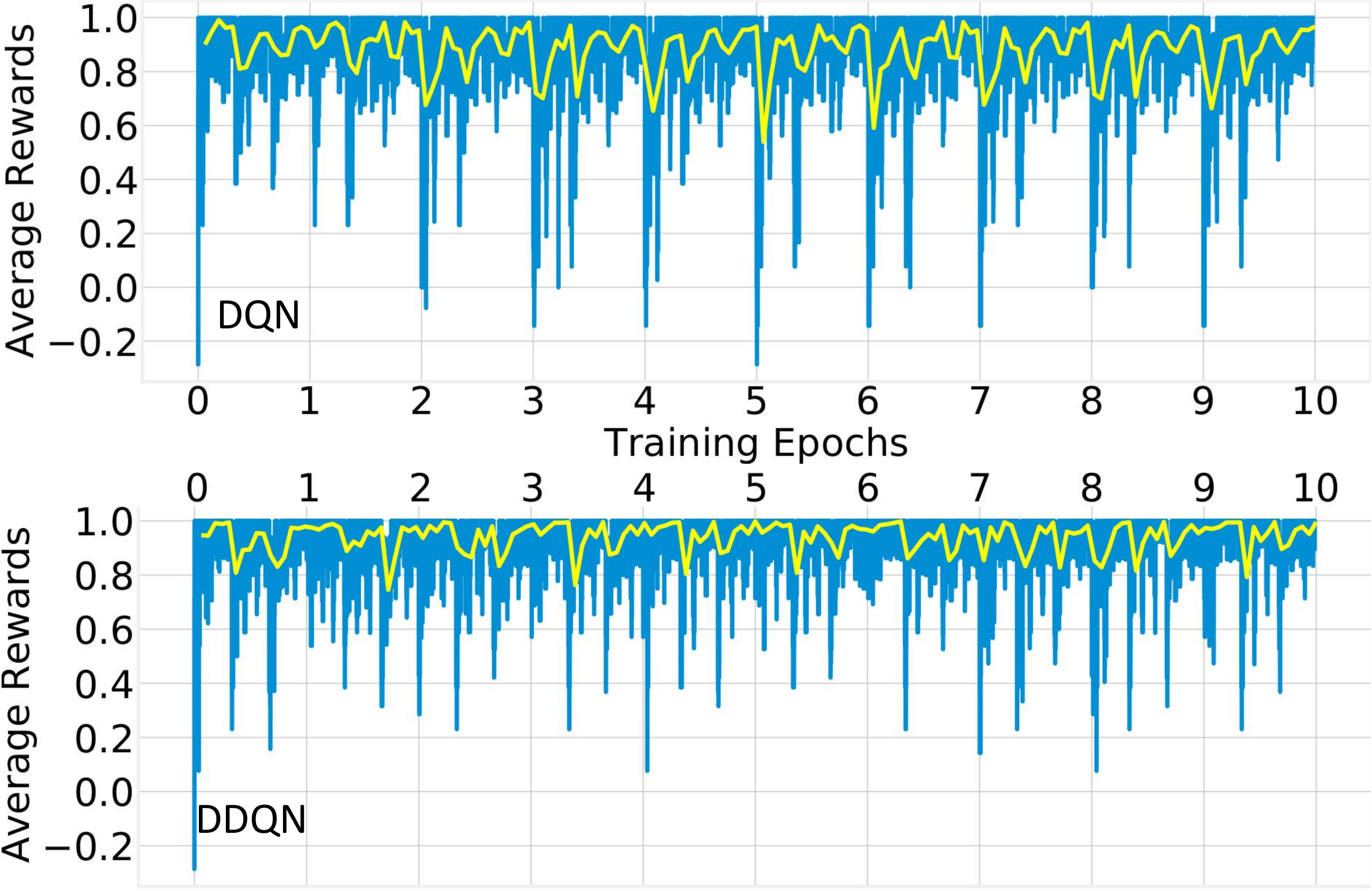}
\caption{Training Results\label{fig:train}}
\end{figure}
\vspace{-3mm}
\par\noindent\textbf{Training Results} We set the number of training runs per epoch i.e. $num\_runs$ to be 100. The training algorithm processes the same oracle request $\mathcal{R}$, i.e. it explores  schedules for the same resultant set of DAGs $\mathcal{G}$ for a total of 100 episodes before processing the next oracle request  $\mathcal{R}^\prime$. Additionally, when the training algorithm moves from processing one request $\mathcal{R}$ to the next request $\mathcal{R}^\prime$, it is ensured that only one period value in the request $\mathcal{R}$ is changed to yield $\mathcal{R}^\prime$. This is done so that during the training phase, after learning the Q-Network for a given oracle request $\mathcal{R}$, the RL environment is not drastically changed during processing of request $\mathcal{R}^\prime$. The neural network architectures used for both the DQN and DDQN contains  1 input layer of size 10, 1 hidden layer of size 16 with ReLU activation and one output layer of size 12 equipped with a softmax function for predicting action probabilities. The corresponding training results are summarized in Fig. \ref{fig:train}.
In both the sub-figures, the x-axis is labelled with the training epoch number where each epoch consists of a total of $81*100=8100$ episodes. Each point of the blue line plot represents the mean reward for each episode. The yellow line plot presents a general trend for the reward where each point represents the mean reward averaged over a consecutive set of 500 episodes. From the two sub-figures, we may conclude that DDQN presents stable training behaviour compared to the DQN where the rewards are oscillating between positive and negative values for the entire duration of training.  This may be attributed to the fact that DQN training performance suffers from over-estimating error values during the training process thereby learning sub-optimal policies in the process \cite{van2016deep}. We next compare the schedules generated by our DDQN with a baseline policy explained as follows.
\begin{figure}[ht]  
	\centering
	\includegraphics[width=.4\textwidth]{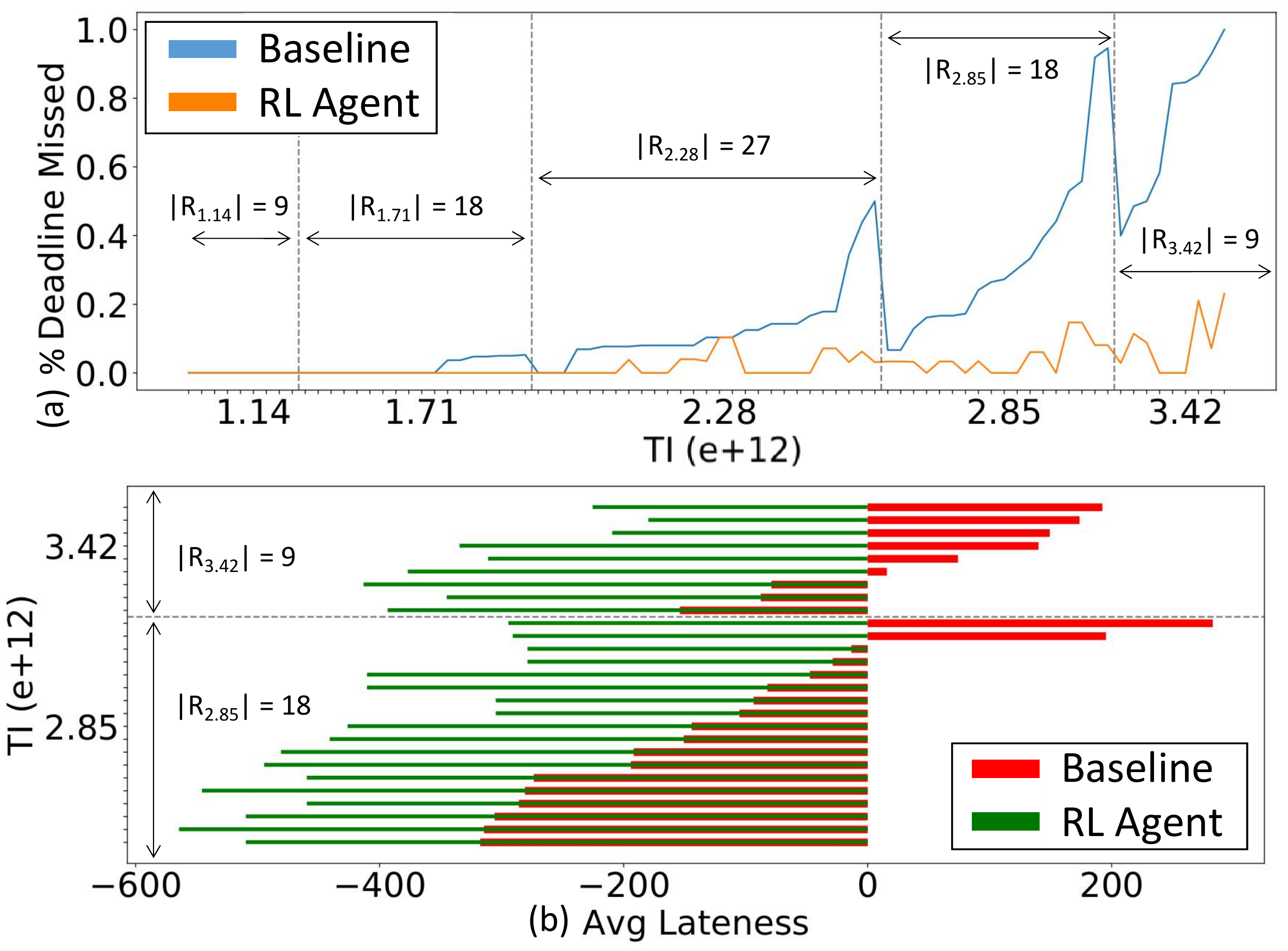}
	\vspace{-2mm}
	\caption{Baseline vs. RL, $|\mathcal{R}_{x}|$ $:\#$ oracle requests with $TI(\mathcal{R})=x$\label{fig:compare}}
\end{figure}
\vspace{-3mm}
\par\noindent\textbf{Testing Results:} We leverage the classical Global EDF scheduling algorithm  outlined in \cite{qamhieh2013global} which is a dynamic EDF scheduling algorithm for executing DAGs on multicore processors as our baseline policy. The algorithm at any point of time considers a task $t_i \in \mathcal{F}$ with the minimum local deadline and simply dispatches it to the device on which the WCET of the task $t$ i.e. $\tau(t_i)$ is minimum. A comparative evaluation between the schedules observed using the baseline algorithm and the schedules reported by our RL scheme is elaborated using Fig. \ref{fig:compare}.
The blue line plot in Fig. \ref{fig:compare}(a), represents percentage of deadline misses observed in a hyper-period for schedules determined by the baseline algorithm and the orange line plot represents the same dictated by the DDQN. Each point on the y-axis represents the deadline miss percentage and each point on the x-axis represents an oracle request $\mathcal{R}$ characterized by a throughput index ($TI$) value. The throughput index represents the throughput requirement for an oracle request $\mathcal{R}$ in terms of Floating Point Operations per second (FLOPs) and is calculated as $TI(\mathcal{R})=\sum_i  FLOPs(G_i)\times w_i/p_i$ where $FLOPs(G_i)$ represents the total number of floating point operations required for a detection pipeline, $w_i$ and $p_i$ are as discussed earlier. It may be observed that for oracle requests demanding higher $TI$ values, the percentage of deadline misses increases upto $80\%$ for the baseline algorithm whereas it remains below $20\%$ for the RL scheme. Fig. \ref{fig:compare}(b) represents a horizontal bar chart where each bar gives the average lateness observed for an oracle request (green bars for RL schemes and red bars for baseline). For an oracle request, the average lateness of the resultant set of DAGs $\mathcal{G}$ is computed by averaging over the individual lateness values of each DAG.  It may be observed that for oracle requests with the higher throughput index, the baseline reports average positive lateness values for most schedules whereas the RL scheme consistently reports negative average lateness values, implying that on the average DAG instances finish before their deadlines using the RL scheme. 
\par \noindent \textbf{Target Platform Results:} We consider that the inference engine admits an oracle request if the deadline miss percentage is less than a threshold $th=15\%$ for the computed schedule. For establishing the efficacy of our low level scheduler, we select borderline admissible schedules (with deadline miss \% $d$ near to but less than  $th$) and present our findings in Table \ref{tab:results}.
Each row represents results for some request $\mathcal{R}$, with the resulting DAG set $\mathcal{G}$ (characterized by  $\langle|\mathcal{G}|, d\rangle$) as reported by the inference engine. In each case, Column 2 provides the number of misses reported by the inference engine, Column 3/4 reports the actual deadline misses when the inferred schedule is deployed  without/with the safe dispatch mode of the low level scheduler being engaged. It may be observed that for schedules suffering from high miss\% under actual deployment without safe mode, the low level scheduler improves their performance with safe mode engaged, thus correcting the deployment badness of the original inference.  
\vspace{-3mm}
\begin{table}[ht]
	\caption{Deployed Low Level Scheduling Results \label{tab:results}}\centering
	\scalebox{0.8}{
	\begin{tabular}{@{}ccll@{}}
		\toprule
		$\langle |\mathcal{G}|,d\% \rangle$ & \begin{tabular}[c]{@{}c@{}}Inference Engine\\ \#misses\end{tabular} & \begin{tabular}[c]{@{}c@{}}Deployed System\\ \#misses\end{tabular} & \begin{tabular}[c]{@{}c@{}}Safe mode\\ \#misses\end{tabular} \\ \midrule
		$\langle 39,10\%   \rangle$        & 4               & 4              & 4                               \\
		$\langle 29,10\%   \rangle$                          & 3                                                                   & 3                                                                  & 1                                                            \\		
		$\langle 35,14\% \rangle$                            & 5                                                                   & 10 $\Rightarrow$ \textbf{28\%} $>\, th$                           & 4   $\Rightarrow$ \textbf{11\%}$<\, th$                                                               \\ \bottomrule
	\end{tabular}
	}
\end{table}
\vspace{-4mm}

\section{Conclusion}
Our proposed combination of RL and low-level  performance recovery technique is possibly the first approach that synergizes AI techniques with real time control-theoretic scheduling techniques towards generating kernel-fusion based runtime mapping decisions for real-time heterogeneous platforms accelerating ADAS workloads.
Future work entails incorporating an edge to cloud feedback mechanism so that on-board mapping decision observations can be leveraged to refine the cloud based inference model using periodic updates.

\bibliographystyle{IEEEtran}
\small
\fontsize{9.0pt}{10.0pt}
\bibliography{reference1.bib}
\selectfont

\end{document}